\documentclass[conference]{IEEEtran}
\IEEEoverridecommandlockouts
\usepackage{cite}
\usepackage{tabularx,booktabs}
\usepackage{multirow}
\usepackage{amsmath,amssymb,amsfonts}
\usepackage{algorithmic}
\usepackage{graphicx}
\usepackage{textcomp}
\usepackage{xcolor}
\usepackage{mathtools}
\newcolumntype{Y}{>{\centering\arraybackslash}X}
\def\BibTeX{{\rm B\kern-.05em{\sc i\kern-.025em b}\kern-.08em
    T\kern-.1667em\lower.7ex\hbox{E}\kern-.125emX}}

\begin{document}

\title{A Dynamic Phasor Framework for Analysis of IBR-Induced SSOs in Multi-Machine Systems \\
\thanks{Financial support from NSF Grant Award ECCS 2317272 is gratefully acknowledged.}
}

\author{\IEEEauthorblockN{Fiaz Hossain, Nilanjan Ray Chaudhuri, and Constantino M. Lagoa}
\IEEEauthorblockA{\textit{School of Electrical Engineering and Computer
Science}, The Pennsylvania State University, University Park, PA, USA \\
emails: fbh5142@psu.edu, nuc88@psu.edu, cml18@psu.edu
}}

\maketitle 

\begin{abstract}
  We propose a generalized dynamic phasor (DP) framework to analyze inverter-based resources (IBRs) connected to multi-machine systems under balanced and unbalanced conditions. It captures subsynchronous oscillations (SSOs) induced by grid-following (GFL) IBRs.  The linearizability and time invariance of the framework enables us to perform eigen decomposition, which is a powerful tool for root-cause analysis of the SSO modes and damping controller design. The same framework also enables analysis of excitation of the SSO modes in presence of data center (DC) loads. The GFL IBRs are modeled in their respective $dq$-frame DPs and the detailed model of synchronous generators (SGs) along with dynamic transmission network models are represented in $pnz$-frame DPs. 
  Several case studies are performed on the modified IEEE two-area benchmark system, where $2$ SGs are replaced by GFL IBRs and validated with EMTDC/PSCAD simulations.  
  First, time- and frequency-domain analyses of the SSO mode are presented followed by the design of a robust decentralized $\mathcal{H}_\infty$ damping controller based on local signals of the GFL IBRs. Second, the dynamic behavior of the system following an unbalanced fault is demonstrated that is damped by the proposed damping controller. Finally, excitation of the SSO mode in presence of DC load is exhibited and its locational impact is analytically quantified.
\end{abstract}

\begin{IEEEkeywords}
Dynamic phasor, subsynchronous oscillation, data center, robust control, DC, IBR, SSO, $\mathcal{H}_\infty$.
\end{IEEEkeywords}

\section{Introduction}


Subsynchronous oscillations (SSOs) induced by grid-following inverter-based resources (GFL IBRs) are becoming prevalent in grids \cite{Real_SSO_event} and recent growth of data center (DC) loads can potentially excite such modes and pose reliability risks \cite{NERC_2025}. Although such phenomena can be captured using the state-of-art commercial electromagnetic transient (EMT) simulation platforms, they have certain limitations including scalability issues due to heavy computational burden and unsuitability in deriving linear time invariant models that allow frequency-domain analysis.

Dynamic phasor (DP)-based modeling approach \cite{Verghese-91-DP} works as a complementary tool that  can employ adaptive time-steps and run faster time-domain simulations compared to traditional EMT platforms because -- (a) the users can select the order of DP coefficients to limit the frequency of modes captured, and (b) the framework is time-invariant. The framework is capable of simulating large disturbances including balanced and unbalanced short circuit faults in presence of IBRs. 
The linearizability and time-invariance of DP models allow eigenvalue decomposition that can help identify the SSO modes,  perform root cause analysis, and facilitate design of controllers, including damping controllers of such modes. 

In published literature, application of the DP framework has been largely confined to systems involving a single machine or a single IBR, see for example \cite{Stankovic_assymetric_SMIB,DP_SSR,hossain2025dynamicphasorframeworkanalysis} that considered unbalance simulation. 
Although \cite{Vega_Herrera} claims to develop a DP-based model of a one-synchronous generator (SG)-two-IBR system, the concepts of DP and space phasor are used interchangeably in that work. Moreover, the modeling framework only considers balanced conditions, and the component-level models are hardly described. The DP-based model of an aircraft power system is presented in \cite{YangTao} where a few SGs are connected in parallel, and unlike terrestrial grids, there are no  transmission lines and IBRs. Therefore, the scalability of the DP-based framework in modeling multi-machine systems with IBRs capable of unbalanced simulation has thus far not been shown. To our knowledge, this is one of the first works that takes a step towards filling that gap.


The main contributions of this paper are the following.\\
1) We present a generalized DP-based modeling framework for multi-machine systems integrated with IBRs that is capable of analyzing balanced and  unbalanced conditions.\\
2) The model captures SSOs induced by GFL IBRs, shows reasonable agreement with EMT simulations, and demonstrates a significant speedup in a modified IEEE benchmark system.\\
3) We show that the time-invariance and linearizability of the framework can be leveraged to design a robust decentralized $\mathcal{H}_\infty$ controller for damping the poorly-damped SSO mode using multiple IBRs as actuators. \\
4) We show that even when the SSO mode is well-damped, it can be excited by DC loads, and the locational impact of the DC load on the excitation can be predicted analytically.


\section{Fundamentals of Dynamic Phasor}
 The generalized averaging theory was first proposed in \cite{Verghese-91-DP}, which suggests that a near-periodic (possibly complex) time-domain waveform $x(\tau)$ in the interval $\tau \in (t - T, t]$ can be expressed using a Fourier series of the form $x(\tau) = \sum_{k = -\infty}^{\infty} X_k(t) e^{jk\omega_s\tau}$, where $\omega_s = \frac{2\pi}{T}$, $k \in \mathbb{Z}$, and $X_k(t)$ are the complex Fourier coefficients that vary with time as the window of width $T$ slides over the signal. The $k$th coefficient, also known as the \textit{$k$th DP}, can be calculated at time $t$ by the following \textit{averaging} operation $X_k(t) = \frac{1}{T}\int_{t-T}^{t} x(\tau) e^{-jk\omega_s\tau}d\tau = \left \langle x \right \rangle_k (t)$. In the DP framework, we are interested in a good approximation provided by the set $\mathcal{U}$ of dominant Fourier coefficients such that $x(\tau) \approx  \sum_{k \in \mathcal{U}} \left \langle x \right \rangle_{k}(t) e^{jk\omega_s\tau}$. Therefore, the generalized averaging-based method leads to an approximated model. From now on, we drop the time variable from DP notations for sake of simplicity. The following are some of the useful properties of DPs: (1) $    \left \langle \frac{\mathrm{d} x}{\mathrm{d} t} \right \rangle_k = \frac{\mathrm{d} \left \langle x \right \rangle_{k}}{\mathrm{d} t} + jk\omega_s\left \langle x \right \rangle_{k}$; (2) if $x(\tau)$ is real, then $    \langle x \rangle_k=\langle x \rangle_{-k}^*$; and (3) in $pnz$ frame,  $\langle x_p \rangle_{-k} = \langle x_n \rangle_k^*$.

\begin{figure}
	\centering
	\includegraphics[width= 0.45\textwidth]{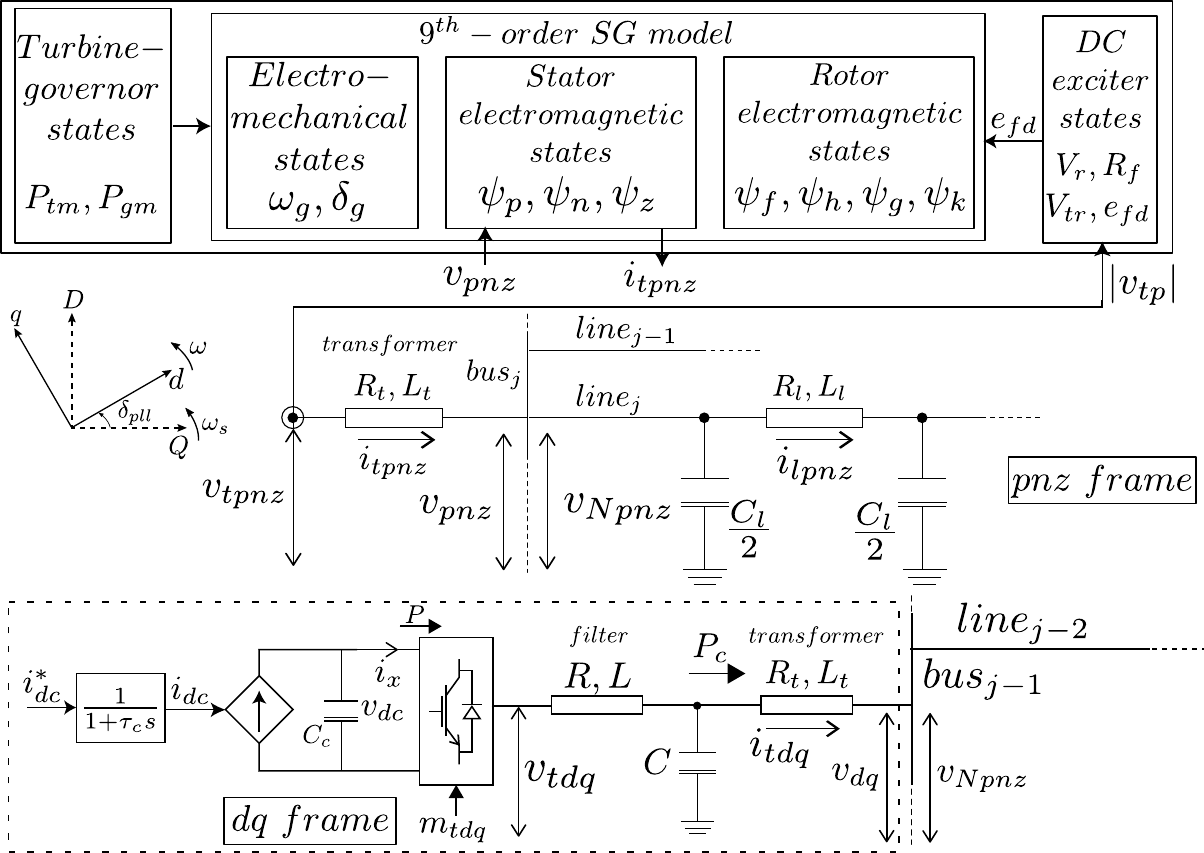}
	 \vspace{-5pt}
	\caption{Proposed genralized DP-based modeling framework capable of analyzing balanced/unbalance conditions. 
    }
	\label{fig:DP_framework}
	\vspace{-10pt}
\end{figure}

Note that the time-domain waveform $x(\tau)$ above can be $abc$ phase quantities, asynchronous $dq0$ frame quantities, or synchronous $DQ0$ frame quantities (note the orientation of axes used in Fig.~\ref{fig:DP_framework}). Using the transformation in synchronously rotating reference frame, the relationship between $DQ0$ and $pnz$ quantities can be obtained in terms of DPs. 
\vspace{-3pt}
\small
\begin{equation}\label{eqn:DQtopnz}
\resizebox{\columnwidth}{!}{$
    \begin{aligned}
        &\langle x_D\rangle_k=\frac{\langle x_p\rangle_{k+1}+\langle x_n\rangle_{k-1}}{\sqrt{2}};~
        \langle x_Q\rangle_k=-\frac{\langle x_p\rangle_{k+1}-\langle x_n\rangle_{k-1}}{\sqrt{2}j}\\
        &\langle x_p\rangle_{k+1}=\frac{\langle x_D\rangle_{k}-j\langle x_Q\rangle_{k}}{\sqrt{2}};~~
        \langle x_n\rangle_{k-1}=\frac{\langle x_D\rangle_{k}+j\langle x_Q\rangle_{k}}{\sqrt{2}};~\langle x_z \rangle_k = \langle x_0 \rangle_k 
    \end{aligned}
    $}
\end{equation}
\normalsize
The following equations describe the relationship between asynchronous $dq$ frame quantities and $pnz$ frame quantities
\vspace{-3pt}
\small
\begin{equation}\label{eqn:pnztodq}
\begin{aligned}
    &\langle x_d \rangle_k = \frac{1}{\sqrt{2}j}\left(e^{j\delta}\langle x_n \rangle_{k-1}-e^{-j\delta}\langle x_p \rangle_{k+1}\right)\\
    &\langle x_q \rangle_k = \frac{1}{\sqrt{2}}\left(e^{j\delta}\langle x_n \rangle_{k-1}+e^{-j\delta}\langle x_p \rangle_{k+1}\right) 
\end{aligned}
\end{equation}
\normalsize
\vspace{-3pt}
where, $\delta$ is the angle between the $Q$-axis and the $d$-axis. For further details on the fundamental concepts, please refer to \cite{demiray2008simulation}.

\section{DP-based Modeling of Multi-Machine System Integrating IBRs}

\subsection{Proposed genralized DP-based modeling framework capable of analyzing balanced/unbalance conditions}
Our proposed framework is shown in Fig.~\ref{fig:DP_framework} that has the following features.

1. \emph{IBR modeling in $dq$-frame:} The controllers of IBRs are typically based on vector control in $dq$ frame.  Hence, it is logical to consider the dominant DP coefficients in the same $dq$-frame for the averaged model of IBR circuit and the controllers instead of representing them in another frame. Usually, zero sequence current is not allowed to pass through IBRs. Therefore, we choose the DP coefficients $k=0$ and $k=\pm2$ that capture positive and negative sequence quantities, respectively. The $dq$ frame associated with a GFL IBR is uniquely determined by its phase locked loop (PLL), whereas the $dq$ frame for a grid-forming (GFM) IBR is determined by its power-frequency droop control \cite{hossain2025dynamicphasorframeworkanalysis}.

2. \emph{Network and SG modeling in $pnz$ domain:} Transmission lines, loads, and SGs are modeled in the $pnz$ domain, wherein $k=\pm1$ DPs should be considered if the frequency of interest is around the synchronous frequency (e.g., for studying SSOs). The mechanical side of SGs is modeled using only $k=0$ DP since the mechanical variables slowly vary. 

3. \emph{Interfacing IBRs with the rest of the network:} The $dq$ frame based DP quantities such as IBR current injections are converted into $pnz$ frame based DPs using \eqref{eqn:pnztodq}. Similarly, IBR terminal voltages in $pnz$ frame can be converted to $dq$ frame quantities.


In this paper, we consider the modified IEEE two-area benchmark system with 50\% GFL IBR penetration and DC loads as shown in Fig. \ref{fig:testSys}. The component models pertaining to this system are presented below. Interested readers can refer to \cite{hossain2025dynamicphasorframeworkanalysis} for GFM IBR modeling in DP framework. 
\begin{figure}
	\centering
	\includegraphics[width= 0.45\textwidth]{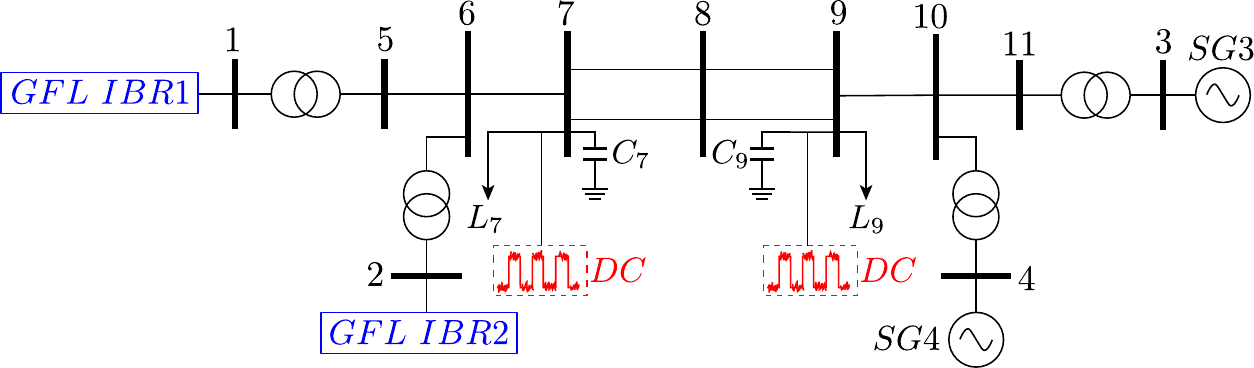}
	 \vspace{-5pt}
	\caption{Modified IEEE two-area benchmark system with 50\% IBR penetration and DC loads.}
	\label{fig:testSys}
	\vspace{-10pt}
\end{figure}
\begin{figure}
	\centering
	\includegraphics[width= 0.45\textwidth]{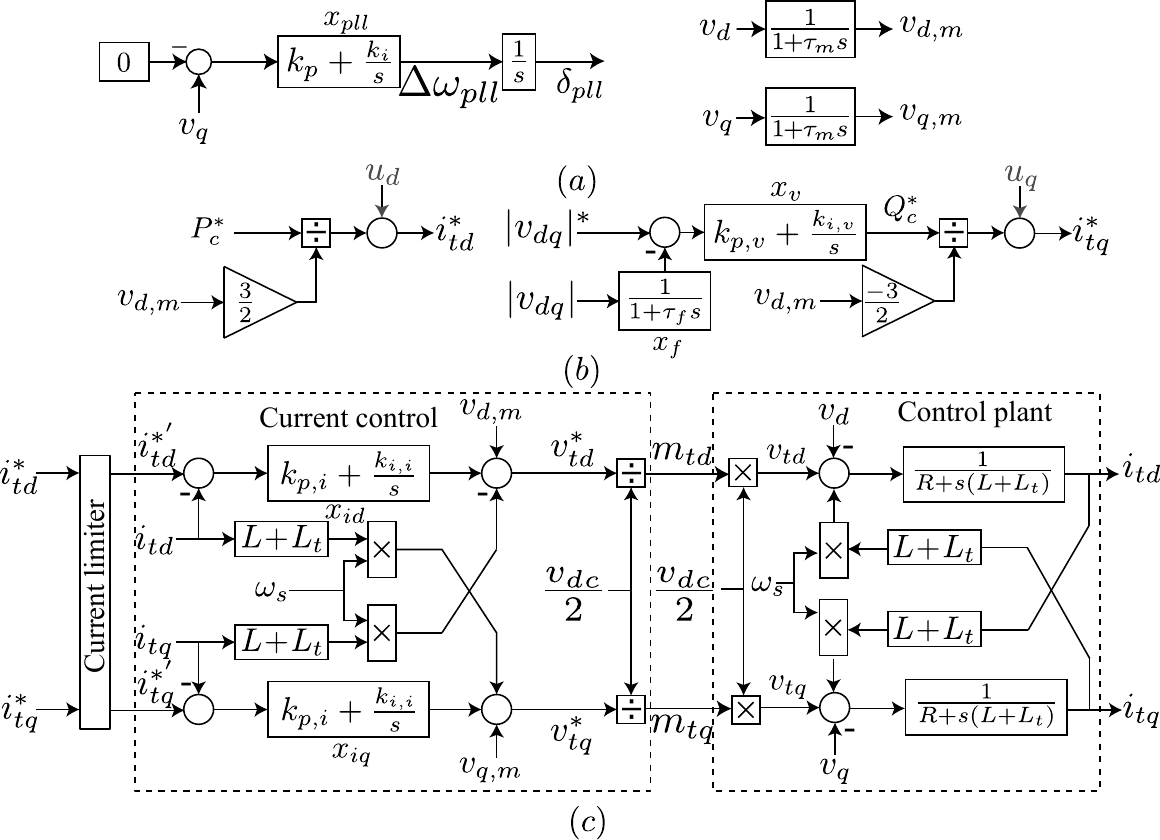}
	 \vspace{-10pt}
	\caption{Block diagram representation of the models of (a) PLL, (b) outer control loops, and (c) inner control loops and control plant. Variables $x$ with subscripts denote the states of the corresponding blocks.}
	\label{fig:GFL IBR_control}
	\vspace{-15pt}
\end{figure}

\subsection{DP model of GFL IBR in $dq$-frame}
Assuming ideal converter, as shown in Fig. \ref{fig:GFL IBR_control}, the GFL IBR model can be divided into the PLL, outer loops regulating real power and voltage, inner current controller, and the plant consisting of the series $R$-$L$ filter (shunt capacitor is not considered) and the transformer impedances shown in Fig.~\ref{fig:DP_framework}. Each component is modeled using $k=0,\pm 2$ DP coefficients. 
\textit{All notations are self-explanatory, and not included due to space constraints.} The PLL (Fig.~\ref{fig:GFL IBR_control}(a)) dynamics shown below determines angle $\delta_{pll}$ between the $Q$ and $d$ axes in Fig.~\ref{fig:DP_framework}.
\small
\begin{equation}
    \begin{aligned}
        &\langle \dot{x}_{pll} \rangle_k = k_{i}\langle v_q \rangle_k-jk\omega_s \langle x_{pll}\rangle_k\\
        &\langle \dot{\delta}_{pll} \rangle_k = k_{p}\langle v_q \rangle_k+\langle x_{pll} \rangle_k-jk\omega_s \langle \delta_{pll} \rangle_k 
    \end{aligned}
\end{equation}
\normalsize
The outer control loop (Fig.~\ref{fig:GFL IBR_control}(b)) models are shown below.
\small
\begin{equation}
    \begin{aligned}
        &\langle \dot{v}_{d,m} \rangle_k = \frac{1}{\tau_{m}}\left(\langle v_d \rangle_k-\langle v_{d,m} \rangle_k\right)-jk\omega_s\langle v_{d,m}\rangle_k\\
        &\langle \dot{v}_{q,m} \rangle_k = \frac{1}{\tau_{m}}\left(\langle v_q \rangle_k-\langle v_{q,m} \rangle_k\right)-jk\omega_s\langle v_{q,m}\rangle_k \\
        &\langle i_{td}^* \rangle_0 = \frac{2}{3}\frac{P_c^*}{v_{d,m}}+\langle u_d \rangle_0,~\langle i_{tq}^* \rangle_0 = -\frac{2}{3}\frac{Q_c^*}{v_{d,m}}+\langle u_q \rangle_0\\
        &\langle \dot{x}_f \rangle_0 = \frac{1}{\tau_f}\left(|v_{dq}|-\langle x_f \rangle_0\right),~\langle \dot{x}_v \rangle_0 = k_{i,v}\left(|v_{dq}|^*-\langle x_f \rangle_0\right)\\
        &Q_c^* = x_v+k_{p,v}\left(|v_{dq}|^*-\langle x_f \rangle_0\right) \\        
    \end{aligned}
\end{equation}
\normalsize
A constant angle current limiter is implemented to limit the current during faults as presented in \cite{hossain2025dynamicphasorframeworkanalysis}. The model of the inner current control loops and the series $R-L$ filter connected to the converter transformer (Fig.~\ref{fig:GFL IBR_control}(c)) is shown below.
\small
\begin{equation}
\resizebox{\columnwidth}{!}{$
    \begin{aligned}
        &\langle \dot{x} _{id} \rangle_k = k_{i,i}\left(\langle i_{td}^{*'} \rangle_k-\langle i_{td} \rangle_k\right)-jk\omega_s \langle x_{id} \rangle_k\\
        &\langle \dot{x}_{iq} \rangle_k = k_{i,i}\left(\langle i_{tq}^{*'} \rangle_k-\langle i_{tq} \rangle_k\right)-jk\omega_s \langle x_{iq} \rangle_k \\
        &\langle v_{td}^* \rangle_k = k_{p,i}\left(\langle i_{td}^{*'} \rangle_k-\langle i_{td} \rangle_k\right)+\langle x_{id} \rangle_k+\langle v_{d,m} \rangle_k-\omega_s \left(L+L_t\right)\langle i_{tq} \rangle_k\\
        &\langle v_{tq}^{*} \rangle_k = k_{p,i}\left(\langle i_{tq}^{*'} \rangle_k-\langle i_{tq} \rangle_k\right)+\langle x_{iq} \rangle_k+\langle v_{q,m} \rangle_k+\omega_s \left(L+L_t\right)\langle i_{td} \rangle_k\\
        &\langle \dot{i}_{td} \rangle_k = \frac{1}{L+L_t}\Big(\langle v_{td} \rangle_k-\langle v_{d} \rangle_k+\omega_s \left(L+L_t\right)\langle i_{tq} \rangle_k\\
        &~~~~~~~~~ -\left(R+R_t\right)\langle i_{td} \rangle_k\Big)-jk\omega_s\langle i_{td} \rangle_k\\
        &\langle \dot{i}_{tq} \rangle_k = \frac{1}{L+L_t}\Big(\langle v_{tq} \rangle_k-\langle v_{q} \rangle_k-\omega_s \left(L+L_t\right)\langle i_{td} \rangle_k\\
        &~~~~~~~~~ -\left(R+R_t\right)\langle i_{tq} \rangle_k\Big)-jk\omega_s\langle i_{tq} \rangle_k
    \end{aligned}
    $}
    \vspace{-10 pt}
\end{equation}
\normalsize
\subsection{DP model of SG in $pnz$ frame}
Except the excitation system, the DP model of SG in the $pnz$ frame is adopted from \cite{demiray2008simulation,Mahipal-thesis}. \textit{The nomenclature of the symbols can be found in these references and is not included in the paper due to space restrictions.} The mechanical side of SGs are modeled considering only the $0$th DP, whereas 
$k=\pm 1$ DPs are considered for modeling the stator transients in $pnz$ domain.  
In addition to the field winding, one damper along the $d$-axis and two dampers along the $q$-axis are modeled, and $k=0,\pm 2$ DPs are considered to model the rotor circuit dynamics. It is worth mentioning that the fluxes and currents of the stator and the rotor are transformed using $\langle f' \rangle_k = \langle f \rangle_k e^{-jk\delta_g}$ in order to make the coupling inductance matrix 
time-invariant.  

The SGs are equipped with IEEE DC1A exciters \cite{DC1A}, which are modeled considering only the $0$th DP. 
\small
\begin{equation}
\resizebox{\columnwidth}{!}{$
    \begin{aligned}
        &\langle \dot{R}_f \rangle_0 = \frac{1}{T_F}\left(\langle e_{fd} \rangle_0-\langle R_f \rangle_0\right),~\langle \dot{V}_{tr} \rangle_0 = \frac{1}{T_r}\left(|v_{p}|-\langle V_{tr} \rangle_0\right)\\
        &\langle \dot{V}_{r} \rangle_0 = \frac{1}{T_A}\{\frac{K_AK_F}{T_F}\left(\langle R_f \rangle_0-\langle e_{fd} \rangle_0\right)+K_A \left(\langle v_{ref} \rangle_0-\langle V_{tr} \rangle_0\right)-\langle V_{r} \rangle_0\}\\
        &\langle \dot{e}_{fd} \rangle_0 = -\frac{1}{T_E}\{K_E \langle e_{fd} \rangle_0+A_{ex}e^{B_{ex}\langle e_{fd} \rangle_0}\langle e_{fd} \rangle_0-\langle V_{r} \rangle_0\}\\
        &\langle v_{f} \rangle_0 = \frac{R_{fd}}{L_{adu}}\langle e_{fd} \rangle_0 
    \end{aligned}
    $}
\end{equation}
\normalsize

\subsection{DP model of transmission network in $pnz$ frame}
In the rest of the paper, unless otherwise specified, the following relationships hold  $\langle x_{pnz}\rangle_k = [\langle x_{p}\rangle_k ~\langle x_{n}\rangle_k ~\langle x_{z}\rangle_k]^T$ and $[A]\langle x_{pnz}\rangle_k$ = $diag(a_p, a_n, a_z)\langle x_{pnz}\rangle_k$. The transmission network considers a lumped $\pi$-section model consisting of the following KCL and KVL algebraic equations.
\small
\begin{equation}
\resizebox{\columnwidth}{!}{$
\begin{aligned}
    &\langle i_{Npnz} \rangle_k = CCI \times [\langle i_{tpnz}\rangle_k^T  ~\langle i_{lpnz}\rangle_k^T]^T,~\langle v_{lpnz} \rangle_k = CCU \times \langle v_{Npnz}\rangle_k
\end{aligned}
$}
\end{equation}
\normalsize
Assuming the network has $n$ nodes, $l$ series $R-L$ branches excluding $m$ SG and $q$ GFL IBR transformers, $\langle i_{Npnz} \rangle_k \in \mathbb{C}^{3n}$ is the vector of net injected currents in the nodes going towards shunt capacitances and any load that may be present, $\langle i_{lpnz}\rangle_k \in \mathbb{C}^{3l}$ and $\langle i_{tpnz}\rangle_k \in \mathbb{C}^{3(m+q)}$ are the vectors of currents flowing through each series $R-L$ branches and transformers, $\langle v_{Npnz}\rangle_k \in \mathbb{C}^{3n}$ is the node voltage vector, $\langle v_{lpnz}\rangle_k \in \mathbb{C}^{3l}$ are the voltages across $R-L$ branches, and $CCI \in \mathbb{R}^{3n \times (3(l+m+q))}$ and $CCU \in \mathbb{R}^{3l \times 3n}$ are the incidence matrix and nodal connectivity matrix, respectively. The transmission network is interfaced with SGs and GFL IBRs according to Fig. \ref{fig:DP_framework}. Only $k=\pm 1$ DPs are considered for modeling the transmission network and loads to capture dynamics around $60$ Hz. 
\small
\begin{equation}
\begin{aligned}
    &\langle \dot{i}_{lpnz} \rangle_k = {\omega_s}[L_l]^{-1}\left(\langle v_{lpnz}\rangle_k-[R_l]\langle i_{lpnz} \rangle_k-jk[L_l]\langle i_{lpnz} \rangle_k\right)\\
    &\langle \dot{v}_{Npnz} \rangle_k = {\omega_s}[C_l]^{-1}\left(\langle i_{cpnz} \rangle_k-jk[C_l]\langle v_{cpnz} \rangle_k\right)
\end{aligned}
\vspace{-10 pt}
\end{equation}
\normalsize
\subsection{DP model of loads in $pnz$ frame}
Constant impedance loads are represented using dynamic models of parallel $R_L-L_L-C_L$ elements at load buses.  
\small
\begin{equation}
    \begin{aligned}
        &\langle \dot{i}_{LLpnz} \rangle_k = {\omega_s}[L_L]^{-1}\left(\langle v_{Npnz}\rangle_k-jk[L_L]\langle i_{LLpnz} \rangle_k\right)\\
        &\langle i_{LRpnz} \rangle_k = [R_L]^{-1}{\langle v_{Npnz}\rangle_k},~\langle i_{Lpnz} \rangle_k  = \langle i_{LLpnz} \rangle_k+\langle i_{LRpnz} \rangle_k
    \end{aligned}
\end{equation}
\normalsize

DC loads are functionally modeled as constant power loads with unity power factor. It is assumed that no negative and zero sequence currents are injected by the DC loads. Since SSO mode excitation due to DC load is studied in this paper, $k=\pm1$ DPs are considered.
\small
\begin{equation}\label{eq:DC_DP}
    \begin{aligned}
        \langle i_{DCp} \rangle_1 = \frac{P_{DC}}{2\langle v_{Np}\rangle_1^{*}},~\langle i_{DCn} \rangle_1=0,~\langle i_{DCz} \rangle_1=0
    \end{aligned}
\end{equation}
\normalsize
\subsection{Fault modeling}
Resistive fault is modeled by the bus fault impedance matrix in $pnz$ frame, which can be expressed as follows.
\small
\begin{equation*}
\resizebox{\columnwidth}{!}{$
\begin{aligned}
    &R_{fabcg} = \begin{bmatrix} R_{fa}+R_g & R_g & R_g\\R_g & R_{fb}+R_g & R_g\\R_g & R_g & R_{fc}+R_g\end{bmatrix},~R_{fpnz} = T^{-1}R_{fabcg}T\\
    &T = \frac{1}{\sqrt{3}}\begin{bmatrix}
         1 & 1 & 1\\
         \alpha^2 & \alpha & 1\\
         \alpha & \alpha^2 & 1
     \end{bmatrix}, ~\alpha=e^{j\frac{2\pi}{3}}
\end{aligned}
$}
\end{equation*}
\normalsize
Note that $R_{fa}$ and $R_g$ are phase $a$-to-neutral (similarly for phases $b$ and $c$), and neutral-to-ground fault resistances, respectively. For example, phase $a$-to-ground fault can be simulated by setting $R_{fa}=R_f,~R_{fb}=\infty,~R_{fc} = \infty$, and $R_g=0$, where $R_f$ is the fault resistance. In practice, very large values are used for $R_{fb}$ and $R_{fc}$. Fault current is integrated in the model using the following equations.
\small
\begin{equation}
    \begin{aligned}
        &\langle i_{fault} \rangle_k = R_{fpnz}^{-1}\langle v_{Npnz} \rangle_k \\
        &\langle i_{cpnz} \rangle_k = \langle i_{Npnz} \rangle_k-\langle i_{Lpnz} \rangle_k-\langle i_{DCpnz} \rangle_k-\langle i_{fault} \rangle_k
    \end{aligned}
\end{equation}
\normalsize

\section{Decentralized SSO Damping Controller Design and Locational Impact of DC Load}

The test system in Fig. \ref{fig:testSys} has a poorly damped SSO mode when the PLLs have a BW of $20$ Hz (Table I). First, we present a robust decentralized $\mathcal{H}_\infty$ controller \cite{pal2005robust} design approach for damping this mode using local signals from IBRs. Next, we propose an  approach to quantify the locational impact of DC load on the excitation of the SSO mode.

\subsection{Robust decentralized damping controller design approach} \label{section: Decentralized control}


\emph{Step 1:} First, the linearized plant model is obtained with respect to an operating point. For each IBR, the corresponding POI voltage magnitude $|v_{dq}|$ can be chosen as the local feedback signal $y$ after passing it through a washout filter, and based on modal controllability, $i_{td}^{*}$ or $i_{tq}^{*}$ can be modulated using control input $u$ (shown as $u_d$ and $u_q$, respectively, in Fig.~\ref{fig:GFL IBR_control}(b)). The disturbance input $w$ can be assumed to be added to $|v_{dq}|$ and the performance signal $z$ can be considered as the difference between $|v_{dq}|^*$ and $|v_{dq}|$ following a filter, $W(s)$. The filter determines the frequency range of signals to be damped. 


\emph{Step 2:} Model order might need to be reduced using approaches such as Schur balanced truncation \cite{pal2005robust} that will be used in the following steps. 

\emph{Step 3:} The $\mathcal{H}_\infty$ controller is designed for one of the IBRs considering the settling time of all modes, $T_s<T_{s1}$. The closed loop system with the controller is reduced as in \textit{Step $2$} and is treated as the new plant for Step $4$.

\emph{Step 4:} The $\mathcal{H}_\infty$ controller is designed for the next IBR considering $T_s<T_{s2}<T_{s1}$. The controller is connected in closed loop, and the process is repeated for all the remaining IBRs until the desired settling time is achieved.

\subsection{Locational impact of DC load in excitation of a mode}

Let, $G(s) = C(sI-A)^{-1}B+D$ be the transfer function of the linearized system model from input $w$ that denotes the real power injection from the DC load to output $y$ where the impact of the DC load is measured. Then $|G_{yw}(j\omega)|$ quantifies the locational impact of the DC load connected to input $w$ towards exciting a mode of angular frequency $\omega$ in the signal $y$. 


\section{Results and Discussions}
The test system in Figure~\ref{fig:testSys} is considered with each IBR and SG producing $\approx700$ MW, and a tie-flow of $400$ MW. The DP model is built in Matlab/Simulink \cite{matlab} and is run using \emph{ode23tb} solver. The EMT model is built in EMTDC/PSCAD \cite{pscad} and is run with $20~\mu s$ time step that considers averaged IBR models and Bergeron models of transmission lines.

\subsection{Comparison between results from DP and EMT models}
\subsubsection{Frequency domain analysis} Table \ref{tab:linearization} shows the comparison of the eigenvalues corresponding to the SSO mode obtained from the models under PLL BWs of $15$ Hz and $20$ Hz. 
We observe slight differences between the results obtained from the DP and EMT models. This is due to the differences in exciter and governor models of the SGs, and how $|v_{dq}|$ in Fig.~\ref{fig:GFL IBR_control}(b) is measured in DP and EMT frameworks. In addition, lumped $\pi$-model is used to represent transmission line in DP instead of Bergeron model. Further, the estimated SSO mode may vary with the change of parameters in Prony analysis.  
\begin{table}
\caption{SSO mode in different modeling frameworks}
\label{tab:linearization}
\centering
\resizebox{0.35\textwidth}{!}{
\begin{tabular}{|c|c|c|c|c|}
\hline
\textbf{PLL BW} & \multicolumn{2}{c|}{\textbf{15 Hz}} & \multicolumn{2}{c|}{\textbf{20 Hz}}\\ \hline
\textbf{Framework} & \textbf{EMT} & \textbf{DP} & \textbf{EMT} & \textbf{DP}\\ \hline
Approach & Prony & Linearization & Prony & Linearization\\ \hline
$f,\ \mathrm{Hz}$ & 5.622 & 5.068 & 6.458 & 6.357\\ \hline
$\zeta,\ \%$ & 12.9 & 16.3 & 0.8 & 0.3\\ \hline

\end{tabular}
}
\vspace{-10pt}
\end{table}

\begin{table}
\caption{Runtime comparison between EMT\textsuperscript{*} and DP models}
\label{tab:runtime}
\centering
\resizebox{0.35\textwidth}{!}{
\begin{tabular}{|c|c|c|c|c|}
\hline
\textbf{Framework} & \textbf{Max (s)} & \textbf{Min (s)} & \textbf{Mean (s)} & \textbf{Std.  dev. (s)}\\ \hline
EMT & 32.397 & 26.772 & 30.309 & 1.178 \\ \hline
DP & 15.454 & 14.453 & 14.763 & 0.222\\ \hline

\end{tabular}
}
\vspace{-5pt}
\small
\begin{flushleft}
\hspace{30pt}\footnotesize \textsuperscript{*} averaged model of IBRs used for fair comparison 
\end{flushleft}
\normalsize
\vspace{-15pt}
\end{table}

\begin{figure}[!b]
	\vspace{-20pt}
	\centering
	\includegraphics[trim = {4cm 9.5cm 4cm 8.5cm}, clip,width= 0.35\textwidth]{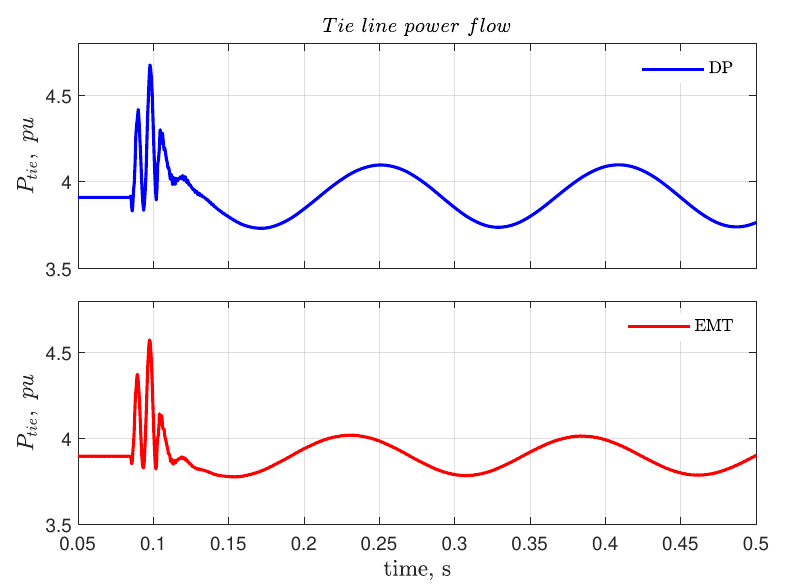}
	 \vspace{-10pt}
	\caption{Comparison of tie line power flow between DP and EMT models following a L-G fault at bus 10. A $20$-Hz PLL BW is considered.}
	\label{fig:LG fault}
	\vspace{-7.5pt}
\end{figure}
\begin{figure}[!b]
	\vspace{-5pt}
	\centering
	\includegraphics[trim = {4cm 9.5cm 4cm 8.5cm}, clip,width= 0.35\textwidth]{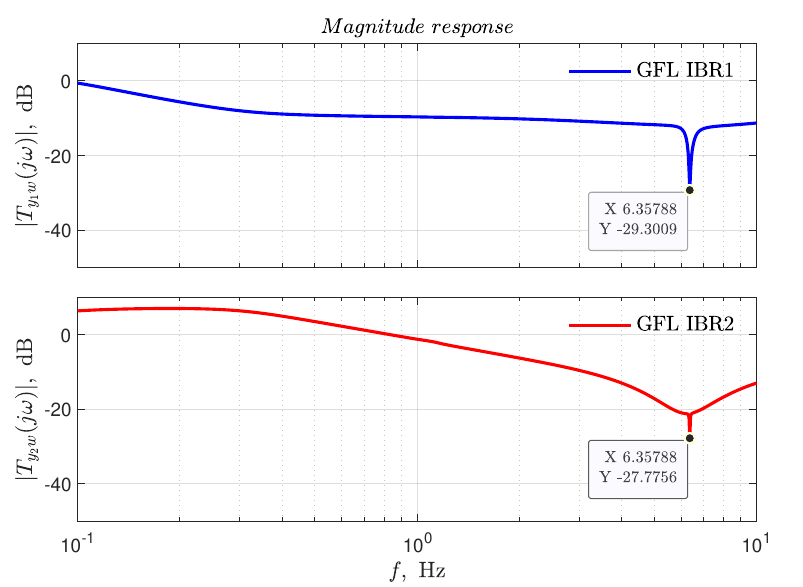}
	 \vspace{-10pt}
	\caption{Frequency response of the transfer functions from perturbation ($w$) to GFL IBRs' individual feedback signals $y_1$ and $y_2$ for 20-Hz PLL BW.}
	\label{fig:Magnitude response}
	\vspace{-15pt}
\end{figure}

\subsubsection{Unbalanced fault simulation}
A single-line-to-ground ($L$-$G$) fault is applied at bus $10$ in Fig.~\ref{fig:testSys}. The fault is applied at $t=0.084~s$ and is self-cleared after one cycle. A PLL BW of $20$ Hz is used. Figure \ref{fig:LG fault} shows the tie-line power flow from bus $7$ to bus $8$ in DP and EMT models. It is found that the DP framework captures the transients during fault and the poorly damped SSO mode persists after the fault clears with slightly higher amplitude compared to the EMT model.

\subsubsection{Simulation speed up in DP} Table II compares the cpu times of EMT and DP models to simulate the unbalanced $L$-$G$ fault mentioned above when the simulations are run for $5$ s after fault clearance. Statistical analysis of $50$ simulations shows that on average the DP model runs $2$x faster.

\subsection{SSO damping controller: design and performance}
\begin{figure}
	\vspace{-5pt}
	\centering
	\includegraphics[trim = {4cm 8.5cm 4cm 8.5cm}, clip,width= 0.35\textwidth]{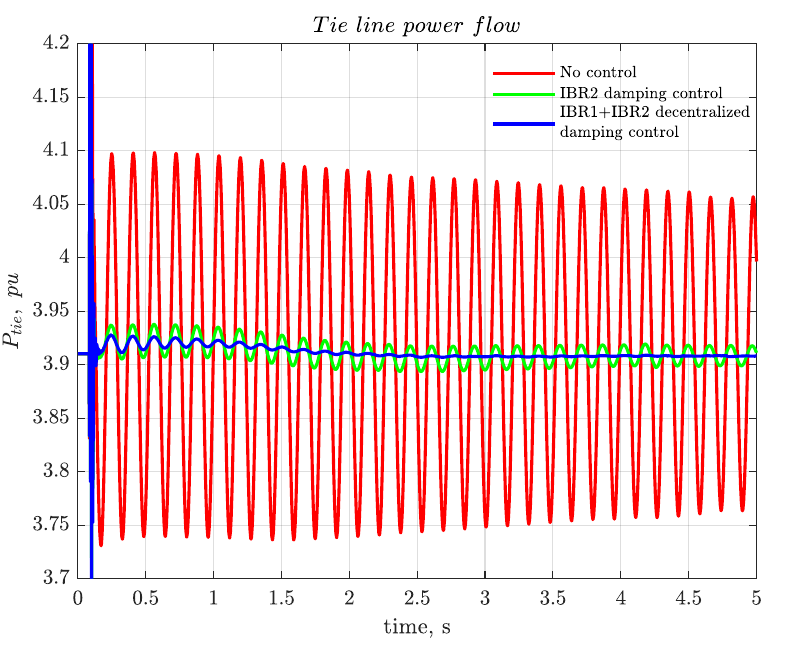}
	 \vspace{-10pt}
	\caption{Performance of decentralized controller for damping SSO mode following L-G fault at bus 10. A $20$-Hz PLL BW is considered.}
	\label{fig:Decentralized control}
	\vspace{-7.5pt}
\end{figure}

For the $20$ Hz PLL BW case, we followed the approach mentioned in subsection \ref{section: Decentralized control}. For both IBRs, the modal controllability of $i_{tq}^*$ was found to be larger, and thus $u_q$ in Fig.~\ref{fig:GFL IBR_control}(b) was chosen. A bandpass filter $W(s) = \frac{25.13s}{s^2+25.13s+1593}$ is used to damp the oscillation with respect to the SSO mode. First, the $\mathcal{H}_\infty$ controller is designed for GFL IBR2 considering $T_s<45~s$ after reducing the plant to the order of $15$ using Schur balanced truncation. Then the second controller is similarly designed for GFL IBR1 considering $T_s<15~s$. 
Figure \ref{fig:Magnitude response} shows the frequency response of the transfer functions from perturbation $w$ to GFL IBRs' individual feedback signals $y_1$ and $y_2$.
Figure \ref{fig:Decentralized control} shows the performance of the IBR2's damping controller and the decentralized controllers of both IBRs compared to the case without control. 

\begin{figure}
	\vspace{-5pt}
	\centering
	\includegraphics[trim = {4cm 9.5cm 4cm 8.5cm}, clip,width= 0.35\textwidth]{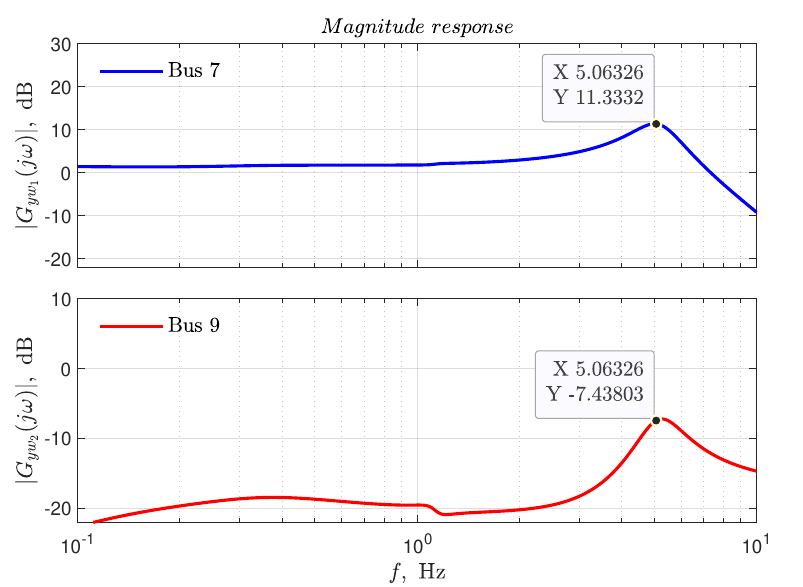}
	 \vspace{-10pt}
	\caption{Magnitude responses of the transfer functions from DC load input (separately at bus $7$ and $9$) to tie-line power output for a $15$-Hz PLL BW.}
	\label{fig:bode_DC}
	\vspace{-7.5pt}
\end{figure}

\begin{figure}[!t]
	\vspace{-5pt}
	\centering
	\includegraphics[trim = {4cm 9.2cm 4cm 8.5cm}, clip,width= 0.35\textwidth]{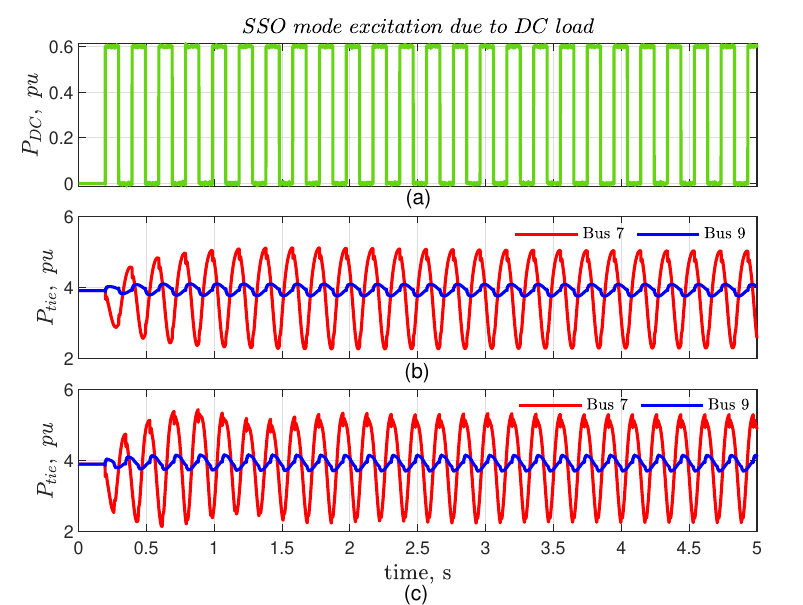}
	 \vspace{-10pt}
	\caption{(a) DC load active power and SSO mode excitation due to DC load at bus 7 and bus 9 in (b) DP, and (c) EMT. PLL BW is 15 Hz.}
	\label{fig:SSO_DC}
	\vspace{-15pt}
\end{figure}

    
    

\subsection{SSO mode excitation due to DC load}
We consider a PLL BW of $15$ Hz for which the SSO mode is well-damped. Bode plots in Fig. \ref{fig:bode_DC} show the locational impact of the DC load in bus $7$ vs bus $9$ on the tie-line power flow. We consider $P_{DC}$ in \eqref{eq:DC_DP} to be a periodic pulse with the frequency of the SSO mode superimposed with band limited white noise as shown in Fig.~\ref{fig:SSO_DC}(a). Figures \ref{fig:SSO_DC}(b) and (c) show that the ratio of amplitudes in tie-flow for the DC load connected to buses $7$ and $9$ is $\approx 6.7$ as observed from both DP and EMT models. The ratio predicted by Fig. \ref{fig:bode_DC} for the SSO mode is $8.7$, which differs slightly from the time-domain result because a periodic pulse consists of infinitely many sine waves. Nevertheless, the proposed index can be quite effective in ranking the locational impact of DC loads on the SSO mode excitation.





\section{Conclusions}
This work takes an important first step towards showing the scalability of the DP-based linearizable time invariant framework in modeling multi-machine systems with IBRs capable of unbalanced simulation. We study a modified IEEE $2$-area benchmark system with $2$ GFL IBRs that demonstrates IBR-induced SSO phenomena. The DP model shows a $2$x speed up on average compared to the EMT model with an averaged IBR representation, when a certain case is simulated $50$ times for $5$ s. We present a robust decentralized $\mathcal{H}_{\infty}$ SSO damping controller design approach using the linearized DP model and demonstrate its effectiveness following an unbalanced fault. Finally, we show that DC loads can excite the IBR-induced SSO mode and its locational impact on the excitation can be quantified using the frequency-dependent gains of the DP-based linearized input-output transfer function models. Our future research will focus on deploying the DP framework for larger systems and designing decentralized controllers using a more systematic approach such as homotopy \cite{homotopy}.

\bibliographystyle{IEEEtran}
\bibliography{Mybib}

\end{document}